\documentclass[11pt]{article}
\usepackage{amsmath,amsthm,amssymb}
\date{}
\usepackage{graphics}
\usepackage{graphicx}
\newtheorem{theorem}{Theorem}
\newtheorem{lemma}[theorem]{Lemma}
\newtheorem{corollary}[theorem]{Corollary}
\newtheorem{definition}[theorem]{Definition}

\begin{document}
\title{The Covering Radius of the Reed--Muller Code $RM(2,7)$ is 40}
\author{  Qichun Wang \footnote{School of Computer Science and Technology, Nanjing Normal University, Nanjing, P.R.China 210046.~E-mail: qcwang@fudan.edu.cn.}}%\and Pantelimon St\u anic\u a \footnote{Department of Applied Mathematics, Naval Postgraduate School, Monterey, CA 93943?216, USA.
% Email: pstanica@nps.edu} }
\maketitle
\begin{abstract}
It was proved by J. Schatz that the covering radius of the second order Reed--Muller code $RM(2, 6)$ is 18 (IEEE Trans Inf Theory 27: 529--530, 1985). However, the covering radius of $RM(2,7)$ has been an open problem for many years. In this paper, we prove that the covering radius of $RM(2,7)$ is 40, which is the same as the covering radius of $RM(2,7)$  in $RM(3,7)$. As a corollary, we also find new upper bounds for $RM(2,n)$, $n=8,9,10$.
\end{abstract}
\par {\bf Keywords:} Reed-Muller codes, covering radius, Boolean functions, second-order nonlinearity.
\par {\bf MSC 2010:} 94B65.

\section{Introduction}
The covering radius of the first order Reed--Muller code $RM(1, n)$ is $2^{n-1}-2^{n/2-1}$ for $n$ even \cite{Rothaus}. For odd $n\le 7$, it equals $2^{n-1}-2^{(n-1)/2}$ \cite{Berlekamp,Hou1,Mykkeltveit}.
However, for odd $n>7$, the covering radius of $RM(1,n)$ is still unknown, although, some bounds have been given~ \cite{Hou2,Hou3,Kavut,Kavut2,Patterson}.

In \cite{Schatz}, Schatz proved that the covering radius of the second order Reed--Muller code $RM(2,6)$ is 18. For $n\geq 7$, the covering radius of $RM(2,n)$ is still unknown. Particularly, the covering radius of $RM(2,7)$  has been an open problem for many years \cite{Carlet1,Carlet2,Cohen,Cohen1,WangS}. In \cite{Hou9}, Hou pointed out that every known covering radius was attained by a coset of $RM(r,n)$ in $RM(r + 1,n)$ and conjectured that the covering radius of $RM(2,7)$ is 40.

For $n\geq 7$, the covering radius of $RM(3,n)$ is also unknown \cite{McLoughlin}. In \cite{WangT}, the authors proved that the covering radius of $RM(3,7)$ in $RM(4,7)$ is 20.

It is also interesting to study the covering radius of the Reed-Muller code in the set of cryptographic Boolean functions (see e.g. \cite{Burgess,Kurosawa}). Particularly, the covering radius of $RM(1,8)$ in the set of balanced Boolean functions is still an open problem.

 In this paper, we prove that the covering radius of $RM(2,7)$ is 40, which is the same as the covering radius of $RM(2,7)$  in $RM(3,7)$ and gives a positive answer to the conjecture proposed by Hou. As a corollary, we also find new upper bounds for $RM(2,n)$, $n=8,9,10$.

\section{Preliminaries}

Let $\mathbb{F}_{2}^{n}$ be the $n$-dimensional vector space over the
finite field $\mathbb{F}_{2}$. We denote by $B_{n}$ the set
of all $n$-variable Boolean functions, from $\mathbb{F}_{2}^{n}$ into $\mathbb{F}_{2}$.

Any Boolean function $f\in B_{n}$ can be uniquely represented
as a multivariate polynomial in
$\mathbb{F}_{2}[x_{1},\cdots,x_{n}]$, called {\em algebraic normal form} (ANF),
\[
f(x_1,\ldots ,x_n)=\sum_{K\subseteq \{1,2,\ldots ,n\}}a_K\prod_{k\in K}x_k,\quad a_K\in\mathbb{F}_2.
\]
The {\em algebraic degree} of $f$, denoted by $\deg(f)$, is the
number of variables in the highest order term with nonzero
coefficient.
 A Boolean function is {\em affine}
if all its ANF terms have degree $\leq 1$. The set of all affine functions is denoted by~$A_{n}$.
 The {\em Hamming weight} of $f$ is the  cardinality of the set $\{x\in \mathbb{F}_{2}^{n}|f(x)=1\}$. The {\em Hamming distance} between two functions
$f$ and $g$ is the Hamming weight of $f+g$, and will be denoted by $d(f,g)$.

 The {\em nonlinearity} of $f\in B_{n}$ is its
distance from the set of all $n$-variable affine functions, that is,
\[
nl(f)=\min_{g\in A_{n}}d(f,g).
\]
The nonlinearity of an $n$-variable Boolean function is bounded above by $2^{n-1}-2^{n/2-1}$ \cite{Carlet0,cs09,Rothaus}.

The {\em $r$-order nonlinearity} of a Boolean function $f$, denoted by $nl_r(f)$, is its
distance from the set of all $n$-variable functions of algebraic degrees at most $r$.

The $r$-th order Reed-Muller code of length $2^n$ is denoted by $RM(r,n)$. Its codewords are the truth tables (output values) of the set of all $n$-variable Boolean functions of degree $\leq r$. The {\em covering radius} of $RM(r,n)$ is defined as
\[
\max_{f\in B_n}d(f,RM(r,n))=\max_{f\in B_n}nl_r(f).
\]

Two $n$-variable Boolean functions $f_1$ and $f_2$ are called affine equivalent modulo $RM(r,n)$ if there
exist $A\in {GL}_n(\mathbb{F}_2)$ and $b\in \mathbb{F}_{2}^{n}$ such that $f_1(x)=f_2(Ax+b)$ modulo $RM(r,n)$.

We use $||$ to denote the concatenation, that is,
\[
(f_1||f_2)(x_1,\ldots,x_n,x_{n+1})=(x_{n+1}+1)f_1(x_1,\ldots,x_n)+x_{n+1}f_2(x_1,\ldots,x_n),
\]
 where $f_1,f_2\in B_n$. We let $|A|$ denote the cardinality of the set $A$.

%\item[$(1)$] $|Fh_{fun_{4}}(16) \bigcap S_1|=47$, $|Fh_{fun_{4}}(16)\bigcap S_2|=256$ and $|Fh_{fun_{4}}(16)\bigcap S_3|=80$;
%\item[$(2)$] $|Fh_{fun_{6}}(16)\bigcap S_1|=43$, $|Fh_{fun_{6}}(16)\bigcap S_2|=156$ and $|Fh_{fun_{6}}(16)\bigcap S_3|=24$;
%\item[$(3)$] $|(g+Fh_{fun_{4}}(26)) \bigcap S_1|=55$, $|(g+Fh_{fun_{4}}(26)) \bigcap S_2|=632$ and $|(g+Fh_{fun_{4}}(26)) %\bigcap S_3|=336$, for any $g\in Fh_{fun_{4}}(26)$;
%\item[$(4)$] $|(g+Fh_{fun_{6}}(26)) \bigcap S_1|=21$, $|(g+Fh_{fun_{6}}(26)) \bigcap S_2|=322$ and $|(g+Fh_{fun_{6}}(26)) \bigcap S_3|=168$, for any $g\in Fh_{fun_{6}}(26)$.
%\item[$(5)$] $|(g+Fh_{fun_{6}}(26)) \bigcap S_1|=7$, for any $g\in Fh_{fun_{6}}(26)$.

\section{The covering radius of the binary Reed-Muller code $RM(2,7)$ is 40}
Let $f\in B_7$. Then it can be written as $f_1||f_2$, where $f_1,f_2\in B_6$. We need to prove that $nl_2(f_1||f_2)\le 40$.
Let $g\in B_6$. It is well known that $nl_2(g)\le 18$, and $g$ is affine equivalent to $g_0=x_1x_2x_3 + x_1x_4x_5 + x_2x_4x_6 + x_3x_5x_6 + x_4x_5x_6$ modulo $RM(2,6)$, if $nl_2(g)=18$. Moreover, $nl(g_0+g_1)\le 22$, for any $g_1\in B_6$ with $\deg(g_1)\le 2$. Therefore, if $f=f_1||f_2$ and $nl_2(f_1)=18$, then 
\[
nl_2(f)\le d(f_2,g_2)+nl(f_1+g_2)\le 18+22=40,
\]
where $g_2$ is a 6-variable Boolean function of degree at most 2 such that $nl_2(f_2)=d(f_2,g_2)$. Similarly, if $f=f_1||f_2$ and $nl_2(f_1)=17$, then we also have $nl_2(f)\le 40$. In fact, we have the following lemma.
\begin{lemma} [Propositions 11 and 14 of \cite{WangS}]
\label{prop1}
Let $f\in B_7$ and $f=f_1||f_2$. If $nl_2(f)>40$, then $15\le nl_2(f_i)\leq16$, for $i=1,2$.
\end{lemma}
The classification of 6-variable Boolean functions under the affine group has been fully studied (see e.g. \cite{Langevin,Maiorana}). It is known that there are exactly 205 affine equivalence classes modulo $RM(2,6)$. Calculating the second-order nonlinearities of these classes, we have the following two lemmas.

\begin{lemma}
\label{obs1}
Let $f\in B_6$. Then $nl_2(f)=16$ if and only if it is affine equivalent to a function with degree $\geq 3$ part among

(1) $fun_1=x_1x_2x_6+x_1x_3x_5+x_2x_3x_4;$

(2) $fun_2=x_1x_2x_3x_4+x_1x_2x_6+x_1x_4x_5+x_2x_3x_5;$

(3) $fun_3=x_1x_2x_3x_4+x_1x_3x_5+x_1x_4x_6+x_2x_3x_5+x_2x_3x_6+x_2x_4x_5;$

(4) $fun_4=x_1x_2x_3x_6+x_1x_2x_4x_5+x_1x_3x_5+x_1x_4x_5+x_1x_4x_6+x_2x_3x_4;$

(5) $fun_5=x_1x_2x_3x_4x_5+x_1x_3x_5+x_1x_4x_6+x_2x_3x_5+x_2x_3x_6+x_2x_4x_5$.
\end{lemma}
\begin{lemma}
\label{obs2}
Let $f\in B_6$. Then $nl_2(f)=15$ if and only if there is a $g\in B_6$ with $\deg(g)\leq 2$ such that $f+g$ is affine equivalent to one of the following functions:

(1) $fun_6=x_1x_2x_3x_4x_5x_6+x_1x_2x_6+x_1x_3x_5+x_2x_3x_4;$

(2) $fun_7=x_1x_2x_3x_4x_5x_6+x_1x_2x_3x_4+x_1x_2x_6+x_1x_4x_5+x_2x_3x_5+x_4x_5;$

(3) $fun_8=x_1x_2x_3x_4x_5x_6+x_1x_2x_3x_4+x_1x_3x_5+x_1x_4x_6+x_2x_3x_5+x_2x_3x_6+x_2x_4x_5;$

(4) $fun_9=x_1x_2x_3x_4x_5x_6+x_1x_2x_3x_6+x_1x_2x_4x_5+x_1x_3x_5+x_1x_4x_5+x_1x_4x_6+x_2x_3x_4+x_4x_6;$

(5) $fun_{10}=x_1x_2x_3x_4x_5x_6+x_1x_2x_3x_4+x_1x_3x_4+x_1x_5x_6+x_2x_3x_4+x_2x_3x_6+x_2x_4x_5+x_3x_4+x_3x_6+x_4x_5$;

(6) $fun_{11}=x_1x_2x_3x_4x_5x_6+x_1x_2x_3x_6+x_1x_2x_4x_5+x_1x_3x_5+x_1x_4x_5+x_1x_4x_6+x_2x_3x_4+x_2x_3x_6+x_2x_4x_5+x_3x_5+x_4x_5+x_4x_6$;

(7) $fun_{12}=x_1x_2x_3x_4x_5x_6+x_2x_3x_4x_5+x_1x_2x_5x_6+x_1x_3x_4x_6+x_1x_2x_4+x_1x_2x_5+x_2x_3x_5+x_3x_4x_5+x_1x_2x_6+x_3x_4x_6$.
\end{lemma}

\begin{definition}
Given $f\in B_n$, we denote by $Fh_f$ the map from $\mathbb{Z}$ to the power set of $B_n$ as follows:
\[
Fh_f(r)=\{g=\sum_{1\leq i<j\leq n}a_{ij}x_ix_j \ | \ a_{ij}\in \mathbb{F}_2 \ and \ nl(f+g)=r\}.
\]
We let $NFh_f:\mathbb{Z}\to \mathbb{Z}$ be the function defined by $ NFh_f(r)=|Fh_f(r)|$. Clearly,  $NFh_f$ is affine invariant and $\sum_{i=0}^{\infty}NFh_f(i)=2^{n(n-1)/2}$.
\end{definition}
It is noted that $0\in Fh_{fun_i}(nl_2(fun_i))$, where $1\le i\le 12$. We calculate the values of $NFh_f$ for those functions in Lemmas 2 and 3, and have the following lemma.
\begin{lemma}
We have
\begin{enumerate}
\item[$(1)$] $NFh_{fun_1}(16)=448$, $NFh_{fun_1}(26)=0$ and $NFh_{fun_1}(28)=64;$
\item[$(2)$] $NFh_{fun_2}(16)=384$, $NFh_{fun_2}(26)=1024$ and $NFh_{fun_2}(28)=0;$
\item[$(3)$] $NFh_{fun_3}(16)=64$ and $NFh_{fun_3}(i)=0$, for $i\geq 26;$
\item[$(4)$] $NFh_{fun_4}(16)=224$, $NFh_{fun_4}(26)=512$ and $NFh_{fun_4}(28)=0;$
\item[$(5)$] $NFh_{fun_5}(16)=272$ and $NFh_{fun_5}(i)=0$, for $i\geq 26$.
\item[$(6)$] $NFh_{fun_6}(15)=112$, $NFh_{fun_6}(25)=0$ and $NFh_{fun_6}(27)=64;$
\item[$(7)$] $NFh_{fun_7}(15)=96$, $NFh_{fun_7}(25)=1024$ and $NFh_{fun_7}(27)=0;$
\item[$(8)$]  $NFh_{fun_8}(15)=16$ and $NFh_{fun_8}(i)=0$, for $i\geq 25;$ 
\item[$(9)$] $NFh_{fun_9}(15)=72$, $NFh_{fun_9}(25)=512$ and $NFh_{fun_9}(27)=0;$
\item[$(10)$]$NFh_{fun_{10}}(15)=72$, $NFh_{fun_{10}}(25)=256$ and $NFh_{fun_{10}}(27)=0;$
\item[$(11)$] $NFh_{fun_{11}}(15)=40$,  $NFh_{fun_{11}}(25)=544$ and $NFh_{fun_{11}}(27)=0$;
\item[$(12)$] $NFh_{fun_{12}}(15)=66$, $NFh_{fun_{12}}(25)=414$ and $NFh_{fun_{12}}(27)=0$.
\end{enumerate}
\end{lemma}
It is well known that there are three affine equivalent classes of $6$-variable homogeneous quadratic Boolean functions, and their nonlinearities could be 16, 24 or 28. We count the number of functions in $Fh_{fun_{i}}$ with the nonlinearity 16, and display the results in the following lemma, where $S_{16}$ denotes the set of  $6$-variable Boolean functions with the nonlinearity 16. 
\begin{lemma} We have
\begin{enumerate}
\item[$(1)$] $|Fh_{fun_{2}}(16) \bigcap S_{16}|=47$ and $|Fh_{fun_{4}}(16)\bigcap S_{16}|=43$;
\item[$(2)$] $|(g_1+Fh_{fun_{1}}(28)) \bigcap S_{16}|=7$, $|(g_2+Fh_{fun_{2}}(26)) \bigcap S_{16}|=55$ and $|(g_4+Fh_{fun_{4}}(26)) \bigcap S_{16}|=21$, for any $g_1\in Fh_{fun_{1}}(28)$ and $g_i\in Fh_{fun_{i}}(26)$, where $i=2,4$.
\item[$(3)$] $|Fh_{fun_{7}}(15) \bigcap S_{16}|=23$, $|Fh_{fun_{9}}(15)\bigcap S_{16}|=15$, $|Fh_{fun_{10}}(15)\bigcap S_{16}|=24$, $|Fh_{fun_{11}}(15)\bigcap S_{16}|=21$ and $|Fh_{fun_{12}}(15)\bigcap S_{16}|=17$;
\item[$(4)$] $|(g_7+Fh_{fun_{7}}(25)) \bigcap S_{16}|=55$, $|(g_9+Fh_{fun_{9}}(25)) \bigcap S_{16}|=21$, $|(g_{10}+Fh_{fun_{10}}(25)) \bigcap S_{16}|=13$,  $|(g_{11}+Fh_{fun_{11}}(25)) \bigcap S_{16}|<30$ and $|(g_{12}+Fh_{fun_{12}}(25)) \bigcap S_{16}|<30$, for any $g_i\in Fh_{fun_{i}}(25)$, where $i\in \{7,9,10,11,12\}$.
\end{enumerate}
\end{lemma}
\begin{lemma} 
\label{thm2}
Let $f\in B_7$ and $f=f_1||f_2$. If $nl_2(f)>40$, then
\[
Fh_{f_i}(k)\subseteq \cup_{m\ge 41-k} Fh_{f_j}(m),
\]
where $i\neq j\in \{1,2\}$.
\end{lemma}
\proof
Let $g\in Fh_{f_i}(k)$. Then $nl(f_i+g)=k$ and there exists an $l_1\in A_6$ such that $d(f_i,g+l_1)=k$. Since
\[
40<nl_2(f)\le d(f_i,g+l_1)+d(f_j,g+l),
\]
for any $l\in A_6$, we have $d(f_j,g+l)\ge 41-k$. That is, $nl(f_j+g)\ge 41-k$,
and the result follows.
\endproof
\begin{lemma} 
\label{thm2}
Let $f\in B_7$ and $f=f_1||f_2$. If $nl_2(f_1)=nl(f_2)=15$, then $nl(f)\le 40$.
\end{lemma}
\proof Suppose $nl_2(f)>40$. Then by Lemma 7, $Fh_{f_i}(15)\subseteq Fh_{f_j}(27)$, where $i\neq j\in \{1,2\}$. Therefore, $NFh_{f_j}(27)\ge NFh_{f_i}(15)>0$. Then by Lemmas 3 and 5, $f_i$ and $f_j$ are affine equivalent to $fun_6$. However,
\[
NFh_{fun_6}(27)=64<NFh_{fun_6}(15)=112,
\]
which is contradictory to $Fh_{f_i}(15)\subseteq Fh_{f_j}(27)$, and the result follows.
\endproof
\begin{lemma} 
\label{thm2}
Let $f\in B_7$ and $f=f_1||f_2$. If $nl_2(f_1)=nl_2(f_2)=16$, then $nl_2(f)\le 40$.
\end{lemma}
\proof
Suppose $nl_2(f)>40$. By Lemma 7, we have $Fh_{f_i}(16)\subseteq \cup_{m\ge 26} Fh_{f_j}(m)$, where $i\neq j\in \{1,2\}$. Then by Lemmas 2 and 5,
$f_1$ and $f_2$ are affine equivalent to $fun_{i_1}$ and $fun_{i_2}$ modulo $RM(2,6)$, where $i_1,i_2\in \{2,4\}$. Therefore, $f$ is affine equivalent to $fun_{i_1}||(fun_{i_2}(Ax+b)+g)$, where $A\in {GL}_6(\mathbb{F}_2)$, $b\in \mathbb{F}_{2}^{6}$ and $g$ is a $6$-variable homogeneous Boolean function of degree 0 or 2. Moreover,
\[
Fh_{fun_{i_1}}(A^{-1}x)(16)\subseteq g(A^{-1}x)+Fh_{fun_{i_2}}(26),
\]
and
\[
 Fh_{fun_{i_2}(Ax)}(16)\subseteq g+Fh_{fun_{i_1}}(26).
\]
{\em Case} 1: $i_1=4$ or $i_2=4$. If $i_1=4$, then $g\in Fh_{fun_{4}}(26)$ (since $0\in Fh_{fun_{i_2}}(16)$) and
\[
 Fh_{fun_{i_2}(Ax)}(16)\subseteq g+Fh_{fun_{4}}(26).
\]
Therefore,
\[
 Fh_{fun_{i_2}(Ax)}(16)\bigcap S_{16}\subseteq (g+Fh_{fun_{6}}(26))\bigcap S_{16}.
\]
By Lemma 6, $|Fh_{fun_{i_2}(Ax)}(16)\bigcap S_{16}|=43$ or $47$, while $|(g+Fh_{fun_{6}}(26))\bigcap S_{16}|=21$, which is a contradiction. Therefore, if $i_1=4$, then $nl_2(f)\le 40$. Similarly, we have $nl_2(f)\le 40$ for $i_2=4$.\\
{\em Case} 2: $i_1=i_2=2$. We have
\[
 Fh_{fun_{2}(Ax)}(16)\bigcap S_{16}\subseteq (g+Fh_{fun_{2}}(26))\bigcap S_{16}.
\]
Let
$Fh_{fun_{2}}(16)\bigcap S_{16}=\{h_1,\ldots,h_{47}\}$ and 
\[
(g+Fh_{fun_{2}}(26))\bigcap S_{16}=\{g+k_1,\ldots,g+k_{55}\},
\]
where $h_i(Ax)=g+k_{i}$, for $i=1,2,\ldots,47$. Then $h_i+h_j$ is affine equivalent to $g+k_{i}+g+k_{j}$. Therefore, if $nl(h_i+h_j)=16$, then $nl(k_{i}+k_{j})=16$.
However, 
\[
|\{h\in \{h_1,\ldots,h_{47}\} \ | \ |(h+\{h_1,\ldots,h_{47}\})\bigcap S_{16})|\ge 13\}|=45,
\]
which is greater than
\[
|\{k\in \{k_1,\ldots,k_{55}\} \ | \ |(k+\{k_1,\ldots,k_{55}\})\bigcap S_{16})|\ge 13\}|=22,
\]
for any $g\in Fh_{fun_{2}}(26)$. This is a contradiction, and the result follows.
\endproof
\begin{lemma} 
\label{thm2}
Let $f\in B_7$ and $f=f_1||f_2$. If $nl_2(f_1)=16$ and $nl_2(f_2)=15$, then $nl_2(f)\le 39$.
\end{lemma}
\proof
Suppose $nl_2(f)\ge 41$. By Lemma 7, we have $Fh_{f_1}(16)\subseteq \cup_{m\ge 25} Fh_{f_2}(m)$ and $Fh_{f_2}(15)\subseteq \cup_{m\ge 26} Fh_{f_1}(m)$. Then by Lemmas 2, 3 and 5, $f_1$ is affine equivalent to $fun_{i_1}$ modulo $RM(2,6)$ and $f_2$ is affine equivalent to $fun_{i_2}$ modulo $RM(2,6)$, where $i_1\in \{1,2,4\}$ and $i_2\in \{7,9,10,11,12\}$.\\
{\em Case} 1: $i_1=1$. We have 
\[
 Fh_{fun_{i_2}(Ax)}(15)\subseteq g+Fh_{fun_{1}}(28).
\]
where $A\in {GL}_6(\mathbb{F}_2)$ and $g\in Fh_{fun_{1}}(28)$.
Therefore, $NFh_{fun_1}(28)=64\ge NFh_{fun_{i_2}}(15)$ and $i_2=11$. However, by Lemma 6,
\[
|Fh_{fun_{11}}(15)\bigcap S_{16}|=21>7=|(g+Fh_{fun_{1}}(28)) \bigcap S_{16}|,
\]
which is a contradiction, and  $nl_2(f)\le 39$.\\
{\em Case} 2: $i_1=2$. We have 
\[
Fh_{fun_{2}}(A^{-1}x)(16)\subseteq g(A^{-1}x)+Fh_{fun_{i_2}}(25).
\]
By Lemma 6, 
\[
|(g(A^{-1}x)+Fh_{fun_{i_2}}(25))\bigcap S_{16}|\ge|Fh_{fun_{2}}(16)\bigcap S_{16}|=47.
\]
Therefore, $i_2=7$. Let
$Fh_{fun_{2}}(16)\bigcap S_{16}=\{h_1,\ldots,h_{47}\}$ and 
\[
(g(A^{-1}x)+Fh_{fun_{7}}(25))\bigcap S_{16}=\{g(A^{-1}x)+k_1,\ldots,g(A^{-1}x)+k_{55}\},
\]
where $h_i(A^{-1}x)=g(A^{-1}x)+k_{i}$, for $i=1,2,\ldots,47$. 
However, 
\[
|\{h\in \{h_1,\ldots,h_{47}\} \ | \ |(h+\{h_1,\ldots,h_{47}\})\bigcap S_{16})|\ge 13\}|=45,
\]
which is greater than
\[
|\{k\in \{k_1,\ldots,k_{55}\} \ | \ |(k+\{k_1,\ldots,k_{55}\})\bigcap S_{16})|\ge 12\}|=22,
\]
for any $g(A^{-1}x)\in Fh_{fun_{7}}(25)$. This is a contradiction, and $nl_2(f)\le 39$.\\
{\em Case} 3: $i_1=4$. We have 
\[
Fh_{fun_{4}}(A^{-1}x)(16)\subseteq g(A^{-1}x)+Fh_{fun_{i_2}}(25).
\]
By Lemma 6, 
\[
|(g(A^{-1}x)+Fh_{fun_{i_2}}(25))\bigcap S_{16}|\ge|Fh_{fun_{4}}(16)\bigcap S_{16}|=43.
\]
Therefore, $i_2=7$. Let
$Fh_{fun_{4}}(16)\bigcap S_{16}=\{h_1,\ldots,h_{43}\}$ and 
\[
(g(A^{-1}x)+Fh_{fun_{7}}(25))\bigcap S_{16}=\{g(A^{-1}x)+k_1,\ldots,g(A^{-1}x)+k_{55}\},
\]
where $h_i(A^{-1}x)=g(A^{-1}x)+k_{i}$, for $i=1,2,\ldots,43$. 
However, 
\[
|\{h\in \{h_1,\ldots,h_{43}\} \ | \ |(h+\{h_1,\ldots,h_{43}\})\bigcap S_{16})|\ge 12\}|=42,
\]
which is greater than
\[
|\{k\in \{k_1,\ldots,k_{55}\} \ | \ |(k+\{k_1,\ldots,k_{55}\})\bigcap S_{16})|\ge 12\}|=22,
\]
for any $g(A^{-1}x)\in Fh_{fun_{7}}(25)$. This is a contradiction, and $nl_2(f)\le 39$.
\endproof
By Lemmas 1, 8, 9 and 10, $nl_2(f)\le 40$ for any $f\in B_7$. Therefore, we have the following theorem.
\begin{theorem} 
\label{thm2}
The covering radius of the Reed--Muller Code $RM(2,7)$ is 40.
\end{theorem}
Let $f_i\in B_i$, where $i=7,8,9$. Then $nl(f_7)\le 56$, $nl(f_8)\le 120$ and $nl(f_9)\le 244$. Therefore, we have the following corollary.
\begin{corollary}
\label{cor1}
The covering radius of $RM(2,n)$ is at most $96, 216, 460$, for $n=8,9,10$ respectively.
\end{corollary}

%We improve the covering radius of $RM(2,n)$, for $n=7,8,9,10$.
In Table~1, we summarize the best known bounds on the covering radius of $RM(2,n)$ \cite{Carlet1,Carlet2,Cohen,Fourquet} for $8\leq n\leq 12$, showing in boldface the contributions of this paper.
\begin{table}[!t]
\caption{The best known bounds on the covering radius of $RM(2,n)$}

 \begin{center}\begin{tabular}{|c|c|c|c|c|c|}
  \hline

$n$ &   8 & 9 &  10 & 11 & 12 \\
  \hline
lower bound   & 84 & 196 & 400  & 848 & 1760 \\
  \hline
upper bound    & \textbf {96} & \textbf {216} & \textbf {460}  & 956 & 1946 \\
  \hline
 \end{tabular}

     \end{center}
     \end{table}
\vspace{0.3cm}

\section{Conclusion}
 In this paper, we prove that the covering radius of $RM(2,7)$ is 40, and find new upper bounds for $RM(2,n)$, $n=8,9,10$.

\section*{Acknowledgment}
The first author would like to thank the financial support from the National Natural Science Foundation of China (Grant 61572189).
\def\refname{References}


\begin{thebibliography}{999}
\bibitem{Berlekamp} E. R. Berlekamp and L. R. Welch, ``Weight distributions of the cosets of the (32, 6) Reed-
Muller code,'' \emph{IEEE Trans. Inform. Theory} 18(1) (1972), 203--207.

\bibitem{Burgess} Y. Borissov, A. Braeken, S. Nikova and B. Preneel, ``On the Covering Radii of Binary Reed-Muller Codes in the Set of Resilient Boolean Functions,'' \emph{IEEE Trans. Inf. Theory} 51:3 (2005), 1182--1189.

\bibitem{Carlet0} C. Carlet, ``Boolean Functions for Cryptography and Error Correcting Codes,'' Chapter of the monography ``Boolean Models and Methods in Mathematics, Computer Science, and Engineering",  Cambridge University Press, pp. 257--397, 2010. Available: http://www-roc.inria.fr/secret/Claude.Carlet/pubs.html.

\bibitem{Carlet1} C. Carlet, ``The complexity of Boolean functions from cryptographic viewpoint,'' 2006. Available: http://dblp.uni-trier.de/db/conf/dagstuhl/P6111.html

\bibitem{Carlet2} C. Carlet and S. Mesnager, ``Improving the upper bounds on the covering radii of binary Reed--Muller codes,'' \emph{IEEE Trans. Inf. Theory} 53:1 (2007), 162--173.

\bibitem{Cohen}  G. Cohen, I. Honkala, S. Litsyn and A. Lobstein, \emph{Covering Codes}, North--Holland, 1997.

\bibitem{Cohen1} G. Cohen, M. Karpovsky, H. Mattson and J. Schatz, ``Covering radius--survey and recent results,'' \emph{IEEE Trans. Inf. Theory} 31:3 (1985), 328--343.

\bibitem{CL92}
 G. Cohen,   S. Litsyn, ``On the covering radius of Reed-Muller codes'', {\em Disc. Math.} 106--107 (1992), 147--155.

\bibitem{cs09} T.~W.~Cusick, P.~St\u anic\u a, \emph{Cryptographic Boolean Functions and Applications} (2nd ed.), Elsevier--Academic Press, 2017.

\bibitem{Fourquet} R. Fourquet and C. Tavernier, ``An improved list decoding algorithm for the second order Reed--Muller codes and its applications,'' \emph{Des. Codes Cryptogr.} 49 (2008), 323--340.

\bibitem{Hou0} X. D. Hou, ``Some results on the covering radii of Reed-Muller codes,'' \emph{IEEE Trans. Inform. Theory}  39(2) (1993), 366--378.

\bibitem{Hou1} X. D. Hou, ``Covering Radius of the Reed--Muller Code $R(1, 7)$ -- A Simpler Proof,'' \emph{J. Comb. Theory, Ser. A} 74(2) (1996), 337--341.

%\bibitem{Fourquet} R. Fourquet and C. Tavernier, ``An improved list decoding algorithm for the second order Reed--Muller codes and its applications,'' \emph{Des. Codes Cryptogr.} 49 (2008), 323--340.

\bibitem{Hou9} X. D. Hou, ``$GL(m,2)$ Acting on $R(r,m)/R(r-1,m)$,'' \emph{Discrete Mathematics}  149 (1996), 99--122.

\bibitem{Hou2} X. D. Hou, ``On the covering radius of $R(1, m)$ in $R(3, m)$,'' \emph{IEEE Trans. Inform. Theory} 42(3) (1996), 1035--1037.

\bibitem{Hou3} X. D. Hou, ``The Covering Radius of $R(1, 9)$ in $R(4, 9)$,'' \emph{Des. Codes Cryptography} 8(3) (1996), 285--292.

\bibitem{Hou} X. D. Hou, ``On the norm and covering radius of the first order Reed--Muller codes,'' \emph{IEEE Trans. Inform. Theory} 43(3) (1997), 1025--1027.

\bibitem{Langevin} P. Langevin, ``Classification of Boolean functions under the affine group,'' Online: http://langevin.univ-tln.fr/project/agl/agl.html

\bibitem{Kavut} S. Kavut, S. Maitra and M. D. Y\"{u}cel, ``Search for Boolean Functions with Excellent Profiles in the Rotation Symmetric Class,'' \emph{IEEE Trans. Inform. Theory} 53(5) (2007), 1743--1751.

\bibitem{Kavut2} S. Kavut, M. D. Y\"{u}cel, ``9-variable Boolean functions with nonlinearity 242 in the generalized rotation symmetric class,'' \emph{Inf. Comput.} 208(4) (2010), 341--350.

\bibitem{Kurosawa} K. Kurosawa, T. Iwata and T. Yoshiwara, ``New covering radius of Reed--Muller codes for $t$-resilient functions,'' \emph{Selected Areas in Cryptography -- SAC} 2001, LNCS 2259, Springer--Verlag, 2001,   pp. 75--86.


\bibitem{Maiorana} J. A. Maiorana, ``A classification of the cosets of the Reed--Muller code R(1,6),'' \emph{Math. Comp.} 57:195 (1991), 403--414.

\bibitem{McLoughlin} A. McLoughlin, ``The covering radius of the $(m-3)$-rd order Reed-
Muller codes and a lower bound on the $(m-4)$-th order Reed--Muller codes,'' \emph{SIAM J. Appl. Math.} 37(2) (1979), 419--4222.

\bibitem{Mykkeltveit} J. J. Mykkeltveit, ``The covering radius of the (128, 8) Reed--Muller code is 56,'' \emph{IEEE Trans. Inform. Theory} 26(3) (1980), 359--362.

\bibitem{Patterson} N. J. Patterson and D. H. Wiedemann, ``The covering radius of the (215, 16) Reed--Muller
code is at least 16276,'' \emph{IEEE Trans. Inform. Theory} 29(3) (1983), 354--356.

\bibitem{Rothaus}  O. S. Rothaus, ``On bent functions,'' \emph{J. Comb. Theory -- Ser. A}  20:3 (1976),   300--305.

\bibitem{Schatz} J. Schatz, ``The second order Reed-Muller code of length 64 has covering radius 18,'' \emph{IEEE Trans. Inf. Theory} 27:4 (1981), 529--530.

\bibitem{WangS} Q. Wang and P.~St\u anic\u a, `New bounds on the covering radius of the second order
Reed--Muller code of length 128,'' \emph{Cryptogr. Commun.} (2018), https://doi.org/10.1007/s12095-018-0289-2

\bibitem{WangT} Q. Wang, C. H. Tan and T. F. Prabowo, ``On the covering radius of the third order Reed–Muller code $RM(3, 7)$,'' \emph{Des. Codes Cryptogr.} 86:1 (2018), 151--159.
\end{thebibliography}
\end{document}